\documentclass[twocolumn,superscriptaddress,showpacs,apl]{revtex4}
\usepackage{epsfig}
\usepackage{delarray}
\usepackage{amsmath}
\usepackage{bm}

\newcommand{\mol}[3]{Mol. Phys. \textbf{#1}, {#2} (#3)}

\newcommand{\nature}[3]{Nature \textbf{#1}, {#2} (#3)}
\newcommand{\jetp}[3]{J. Exp. Theor. Phys. \textbf{#1}, {#2} (#3)}

\newcommand{\be}{\begin{equation}}
\newcommand{\en}{\end{equation}}
\newcommand{\bega}{\begin{eqnarray}}
\newcommand{\eda}{\end{eqnarray}}

\begin{document}
\title{Gigantic Enhancement of Magneto-Chiral Effect
in Photonic Crystals}

\author{Kei Sawada}
\email{sawada@appi.t.u-tokyo.ac.jp}
\affiliation{Department of Applied Physics, the University of Tokyo, 
7-3-1, Hongo, Bunkyo-ku, Tokyo 113-8656, Japan}

\author{Naoto Nagaosa}
\affiliation{Department of Applied Physics, the University of Tokyo, 
7-3-1, Hongo, Bunkyo-ku, Tokyo 113-8656, Japan}
\affiliation{Correlated Electron Research Center (CERC), 
National Institute of Advanced Industrial Science and Technology (AIST), 
Tsukuba Central 4, Tsukuba 305-8562, Japan}
\affiliation{CREST, 
Japan Science and Technology Agency (JST), Japan}

\date{\today}

\begin{abstract}
We theoretically propose a method to enhance dramatically a magneto-chiral(MC)
effect by using the photonic crystals composed of a multiferroic material. 
The MC effect, the directional birefringence even for unpolarized light, 
is so small that it has been difficult to observe experimentally.
Two kinds of periodic structures are investigated; 
(a) a multilayer and (b) a stripe 
composed of a magneto-chiral material and air. 
In both cases, the difference in reflectivity between different 
magnetization directions 
is enhanced by a factor of hundreds compared with a bulk material.
\end{abstract}

\maketitle


Optical effects in media with broken symmetry
 have been of great interest 
from both scientific and applicational viewpoints.
Broken inversion symmetry ${\cal I}$ 
and time-reversal symmetry ${\cal T}$ give rise to 
polarization-dependent properties; natural optical activity and 
magneto-optical effects, respectively. 
Polarization-independent properties require breaking of both 
${\cal I}$ and ${\cal T}$.
When both ${\cal I}$ and ${\cal T}$ are simultaneously broken, 
we have directional birefringence even for unpolarized light, 
which is non-reciprocal. 
Such directional birefringence is conventionally called 
a magneto-chiral (MC) effect$^1$ 
or an optical magnetoelectric(OME) effect$^2$, 
if the microscopic structure of the medium is helical or polar, respectively. 
Nevertheless, as the polarization-independent non-reciprocal effects, 
the MC effect and OME effect are quite similar, 
and we call both the effects  ``MC effects" for simplicity. 
The MC effect has been mostly studied in chiral media 
in external fields$^{1, 3}$. 
On the other hand, only recently has it been investigated in 
condensed materials without external fields. 
Such materials should be multiferroic, i.e., 
without ${\cal I}$ and ${\cal T}$. 
Multiferroic materials such as 
GaFeO$_3$ exhibit the spontaneous MC effect even in the absence of 
the external fields$^{2, 4, 5}$. 
Optical properties in multiferroics can be described by 
a toroidal moment defined by 
$\vec{T}\equiv \sum_i \vec{r}_i \times \vec{S}_i 
\simeq \vec{P} \times \vec{M}$, where 
$\vec{r}_i$ is the displacement from the center position 
of atoms, $\vec{S}_i$ is a magnetic moment at $i$-th site, 
$\vec{P}$ is the spontaneous electric polarization, 
and $\vec{M}$ is the magnetization, respectively$^6$. 
The dielectric function depends on 
whether light propagation is parallel or antiparallel to the toroidal moment.

Unfortunately the MC effect is usually too small to be observed 
experimentally. 
In this letter we theoretically propose a method to 
enhance the MC effect by using a photonic crystal.
A photonic crystal is a periodic array of dielectric materials$^{7, 8, 9}$. 
Light propagation in a photonic crystal described by a wave equation is 
analogous to electron motion in a solid. 
In this analogy the wave equation with a spatially modulated dielectric 
constant corresponds to the Schr\"odinger equation with a potential. 
However, most studies on photonic crystals do not utilize this analogy 
to electronic systems; 
instead they aim only to realize the localization of 
light or high $Q$-values. 
In condensed materials, magnetism has been one of the central issues, 
while much less attention has been paid to magnetic photonic crystals$^{10}$. 
In  light of rich physics and applications arising from 
magnetism in condensed materials, magnetic photonic crystals should 
have potential importance both for fundamental physics and for application. 
Thus it is highly desirable to construct a theory 
describing light propagation in materials without time-reversal symmetry.

We theoretically found that the difference in reflectivity 
for opposite directions of the toroidal moment is 
magnified by hundreds of times in photonic crystals than that in a bulk.
The enhancement is due to the 
Bragg reflection and a photonic band gap, which are absent in 
a bulk or a Fabry-Perot cavity.
In this respect photonic crystals are of great promise for 
better interferometers 
in place of Fabry-Perot cavities.

We solve the following wave equation derived from Maxwell equations, 
$
\frac{1}{\varepsilon(\vec{r})} \nabla \times
\bigl[ \nabla \times \vec{E} (\vec{r}) \bigr]
=(\frac{\omega}{c})^2 \vec{E} (\vec{r}), 
$
where $\varepsilon(\vec{r})$ is a dielectric function 
and $\vec{E}$ is an electric field vector. 
The direction of the toroidal moment is postulated to be 
parallel to the $x$ axis, 
and the MC effect can be introduced in the 
dielectric function of the medium 1 as 
$
\varepsilon_1=\varepsilon_1^0 +\alpha {\hat k}_x,
$
where $\varepsilon_1^0$ is the static dielectric constant, 
$\alpha$ is a strength of the MC effect 
and ${\hat k}_x$ is an operator representing 
the $x$ component of the unit wavevector. 
We set the dielectric constants of the two media as 
$\varepsilon_1^0=15, \alpha=10^{-6}$ and $\varepsilon_2=1$(air), 
and consider the normal incidence throughout this letter.
In order to solve the equation above, 
we use a transfer matrix method for multilayer structures 
shown in Fig.~1(a), 
and a plane-wave expansion method$^{11}$ for stripe structures 
shown in Fig.~(b).
The polarization change, i.e., magneto-optical effects and 
natural optical activity,
 are neglected throughout this letter, 
and we only focus on the MC effect, which is polarization-independent.
We calculate frequency dependence of the difference in reflectivity 
for opposite directions of the toroidal moments. 
It should be noted that the difference is zero in a conventional material, 
and non-zero only in a magneto-chiral medium.
First of all, we calculate reflectivity without the periodic modulation 
of $\varepsilon(\vec{r})$.
Let us consider light reflection of a magneto-chiral material
with the normal incidence; 
$R_{\rm bulk}^\uparrow -R_{\rm bulk}^\downarrow
= \Bigl| \frac{\sqrt{15+10^{-6}}-1}{\sqrt{15+10^{-6}}+1} \Bigr| ^2
-\Bigl| \frac{\sqrt{15-10^{-6}}-1}{\sqrt{15-10^{-6}}+1} \Bigr| ^2 
\sim 10^{-8}$, where the indices $\uparrow$ and $\downarrow$ denote 
whether $\vec{T}$ is parallel or antiparallel to the $x$-axis.
This value should be compared with that in a periodic structure.
A MC effect in a material is originated from 
the spin-orbit interaction that is nothing but a relativistic effect, 
and is difficult to control microscopically.
In contrast, we can enhance the effect by the artificial periodic 
modulation with the size of the order of light wavelength. \\

(a) {\it Magneto-chiral multilayers}\ \\

The multilayered structure (16 layers) composed of the MC medium and air 
is shown in Fig. 1(a). 
The ratio of the thickness between the MC medium and air is 
put to be 0.2 : 0.8.
We solve the wave equation by means of the transfer matrix method. 
The boundary conditions and the phase change through the propagation 
can be described as matrices, and we can solve the equation simply by 
multiplying the matrices.

We calculate reflectivity $R^{\uparrow, \downarrow}$ where 
the arrows represent the directions of $\vec{T}$ whether it is parallel or 
antiparallel to the $x$ axis. 
Figure 2 shows the frequency dependence of the difference in 
reflectivities for opposite directions of $\vec{T}$. 
Here we introduced small absorption. 
There are two distinct behaviors in the figure.
One is the region characterized by ``gap" in the figure corresponds to 
the photonic band gap.
The other is the region where $R^\uparrow -R^\downarrow$ oscillates, 
corresponding to 
a Fabry-Perot cavity 
that has nothing to do with the periodic structure.
At the edge of the gap, we have a remarkable enhancement.
This enhancement is interpreted as follows.
The different toroidal moment direction has 
the different refractive index, 
leading to different reflectivity. 
Hence the band gaps for the opposite toroidal moment directions 
are at different frequency regions. 
At the gap edge, it can happen that 
the frequency is in the band gap for $\vec{T} \parallel \vec{e}_x$, 
and in the conduction band for $\vec{T} \parallel -\vec{e}_x$, 
because we have the nonreciprocal dispersion 
$\omega (\vec{k})\neq \omega (-\vec{k})$. 
Therefore the difference in reflectivities between 
the two toroidal moment directions is significantly magnified. 
In other words, this difference in reflectivities behaves 
qualitatively similar to $\frac{\partial R}{\partial \omega}$, 
and the differential is singular at the edge of the gap, 
resulting in the enhancement. 
The magnification at the gap edge is robust under small changes of 
incident angles and toroidal moment direction, 
whereas the values in the Fabry-Perot region 
are fragile against these changes. 
We note that if there is no absorption in a MC medium, 
the value of $R^\uparrow -R^\downarrow$ at the gap edge 
is the order of $10^{-5}$, an order of magnitude larger.

\ \\

(b) {\it Magneto-chiral stripes}\\

In the stripe structures shown in Fig. 1(b), 
reflectivity changes as the direction of the toroidal moment is reversed. 
We calculate $R_1^\uparrow -R_1^\downarrow$, 
which is the difference between reflectivities of 
the first-order Bragg reflected waves, denoted by the subscript 1. 
We solve the equation by Fourier transformation in the region $0< z < L$, 
\begin{equation}
E=\displaystyle \sum_{n=0, \pm 1} \sum_{m=0}^\infty
 A_{nm}e^{ik_nx}e^{i\frac{m\pi}{L}z},
\end{equation}
where $k_n=\frac{2\pi n}{a}$, $L$ is the thickness of the stripe, 
and we have a sinusoidal modulation 
$
\frac{1}{\varepsilon(x)}=
\frac{1}{2}( {\hat \varepsilon}_1^{-1}+ \varepsilon_2^{-1} )
+\frac{1}{2}( {\hat \varepsilon}_1^{-1} - \varepsilon_2^{-1})
\cos \frac{2\pi x}{a}, $
where $a$ a periodicity along the  $x$ direction. 
It should be noted that the structure shown in Fig.~1(b) is 
only a schematic one. 
In the calculation, we use the sinusoidal dielectric modulation 
instead of the rectangular one, because the higher order 
Fourier components give no essential contribution to the results.
We choose the dielectric constant of the substrate as 
$\varepsilon_3=15$.
We take 606 Fourier components, $n=-1, 0, 1$, $m=0$-200, 
which are enough for convergence.
Figure 3 represents the frequency dependence of 
the difference in intensities of Bragg reflected waves 
for $L=0.5a, a$, and $2a$.
In the region $\frac{\omega a}{2\pi c} \leq 1.0$, 
the wavenumber is imaginary, implying that the reflected light is 
an evanescent wave. 
Therefore we should focus on the region 
$\frac{\omega a}{2\pi c} \geq 1.0$. 
In the region just above $\frac{\omega a}{2\pi c} = 1.0$,
the reflection angle is close to $90^\circ$, and 
it is experimentally difficult to detect the beam. 
Let us consider the case $L=a$ at first.
We find an enhancement at $\frac{\omega a}{2\pi c} \simeq 1. 3$  
corresponding to the Bragg reflection angle $\theta =63^\circ$, 
which can be observed experimentally.
Moreover, even away from $\frac{\omega a}{2\pi c} \simeq 1.3$,
$R_1^\uparrow -R_1^\downarrow$ is of the order of 
$10^{-6}$, which is two orders of magnitude larger than that in a bulk.
Although there is no band gap in this system, 
the optical effect is remarkably enhanced when 
the wavelength of light is comparable to the periodicity of the 
photonic crystals.
When the thickness of the stripe changes, $L=0.5a$ and $2a$, 
the order of $R_1^\uparrow -R_1^\downarrow$ remains the same. 
It is because 
this enhancement is due to the periodicity along the $x$ direction 
and has nothing to do with the value of the thickness.\\

Let us discuss the relation and the difference between (a) and (b) in Fig. 1. 
In the MC multilayers, 
a periodic structure and the non-reciprocal dispersion 
$\omega (\vec{k})\neq \omega (-\vec{k})$
play essential roles.
On the other hand, in the MC stripes without any band gap, 
a periodic structure brings about multiple scatterings, 
and only the waves that satisfy the Bragg conditions can be diffracted.
In both cases, $\tilde{R}^\uparrow -\tilde{R}^\downarrow$ 
is roughly a differential of $\tilde{R}$ with respect to $\omega$, 
where $\tilde{R}^\uparrow$ is $R^\uparrow$ or $R_1^\uparrow$.
The multiple scatterings induce the oscillation of $\tilde{R}$ and 
the large $\frac{\partial \tilde{R}}{\partial \omega}$, 
resulting in a remarkable enhancement of
 $\tilde{R}^\uparrow -\tilde{R}^\downarrow$.
In order to magnify this effect, 
it is necessary to array the materials 
parallel to the toroidal moment direction.
It should be noted that the enhancement factor of hundreds 
is not specific to these cases ($\alpha =10^{-6}$) 
but valid for another value of $\alpha$.

In conclusion, we have investigated an enhancement of 
the magneto-chiral effect in photonic crystals.
The difference in intensities of two waves 
for different toroidal moment directions
is hundreds of times enhanced from that in a bulk.
The direction of the toroidal moment can be reversed by 
changing the directions of the polarization or the magnetization 
by external fields. 
Our theory is in principle applicable for a wide range of materials, 
by changing dielectric constants and thickness of the layers 
at one's disposal.

The authors are grateful to S. Murakami, N. Kida, Y. Okimoto 
and Y. Tokura for fruitful discussions. 
This work is financially supported by a NAREGI Grant, Grant-in-Aids 
from the Ministry of Education, Culture, Sports, Science and Technology 
of Japan.

\begin{figure}[t]
  \centerline{
    \epsfxsize=6.5cm
    \epsfbox{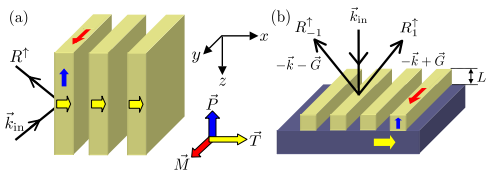}
  }
 \label{layer}
\caption{Schematic illustration of (a) a magneto-chiral multilayer and (b) a stripe structure whose constituent materials are multiferroics and air. 
The vector $\vec{G}$ in (b) denotes a reciprocal lattice vector.}
\end{figure}
\begin{figure}[h]
  \centerline{
    \epsfxsize=5.5cm
    \epsfbox{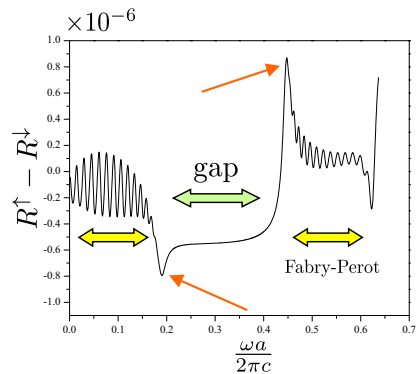}
  }
 \label{layer-chiral}
\caption{Frequency dependence of $R^\uparrow -R^\downarrow$ 
for the normal incidence. 
The small absorption is introduced as $\varepsilon_1^0=15+0.1i$.}
\end{figure}
\vspace{-0.5cm}
\begin{figure}[h]
  \centerline{
    \epsfxsize=5.5cm
    \epsfbox{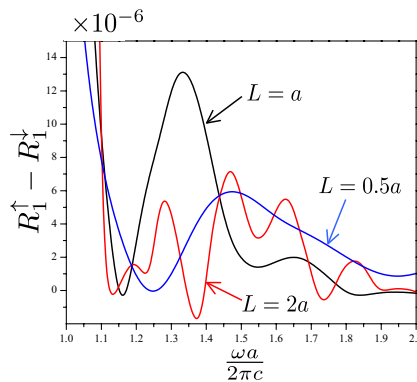}
  }
 \label{stripe-chiral}
\caption{Frequency dependence of $R_1^\uparrow -R_1^\downarrow$
for the normal incidence.}
\end{figure}


\vspace{6cm}

\begin{thebibliography}{99}
\bibitem{barron} L. D. Barron and J. Vrbancich, \mol{51}{715}{1984}.
\bibitem{kubota} M. Kubota, T. Arima, Y. Kaneko, J. P. He, X. Z. Yu and Y. Tokura, 
\prl{92}{137401}{2004}.
\bibitem{rikken} G. L. J. A. Rikken and E. Raupach, \nature{390}{493}{1997}; 
\pre{58}{5081}{1998}; G. L. J. A. Rikken, C. Strohm and P. Wyder, 
\prl{89}{133005}{2002}; C. Koerdt, G. D\"uchs and G. L. J. A. Rikken, 
\textit{ibid}. \textbf{91}, 073902 (2003); 
F. A. Pinheiro and B. A. van Tiggelen, 
J. Opt. Soc. Am. A \textbf{20}, 99 (2003).
\bibitem{jung} J. H. Jung, M. Matsubara, T. Arima, J. P. He, Y. Kaneko 
and Y. Tokura, \prl{93}{037403}{2004}.
\bibitem{murakami} S. Murakami, R. Shindou, N. Nagaosa and A. S. Mishchenko, 
\prb{66}{184405}{2002}.
\bibitem{popov} Yu. F. Popov, A. M. Kadomtseva, G. P. Vorob'ev, V. A. Timofeeva, D. M. Ustinin, A. K. Zvezdin and M. M. Tegeranchi, \jetp{87}{146}{1998}.
\bibitem{yablo} E. Yablonovitch, \prl{58}{2059}{1987}; 
E. Yablonovitch and T. J. Gmitter, \textit{ibid}. \textbf{63}, 1950 (1989); 
E. Yablonovitch, T. J. Gmitter and K. M. Leung, \textit{ibid}. \textbf{67}, 
2295 (1991).
\bibitem{joan} J. D. Joannopoulos, R. D. Meade and J. N. Winn, 
 \textit{Photonic Crystals}, (Princeton University Press, Princeton, 1995).
\bibitem{sakoda1} K. Sakoda, \textit{Optical Properties of 
Photonic Crystals}, (Springer Series in Optical Sciences, Springer, 2001).
\bibitem{haldane} F. D. M. Haldane and S. Raghu, cond-mat/0503588.
\bibitem{sakoda2} K. Sakoda, \prb{52}{8992}{1995}.
\end{thebibliography}
\end{document}